\documentclass[aps,twocolumn,showpacs]{revtex4}

\usepackage{exscale}
\usepackage{graphicx}
\usepackage{amsmath}
\usepackage{latexsym}
\usepackage{amsfonts}
\usepackage{amssymb}
\usepackage{times}

\usepackage{amsmath}
\usepackage{amscd}
\usepackage{latexsym}
\usepackage{bbm}
\usepackage{dcolumn}      
\usepackage{bm}           
\usepackage{pstricks,pst-grad,fancybox,graphics}
\usepackage{amsfonts}
\usepackage{enumerate}

\newcommand{\C}{\ensuremath{\mathbbm C}}
\newcommand{\R}{\ensuremath{\mathbbm R}}

\newcommand{\ket}[1]{\ensuremath{|#1\rangle}}

\newcommand{\ketbra}[1]{\ensuremath{| #1 \rangle \langle #1 |}}

\newcommand{\Eins}{\ensuremath{\mathbbm 1}}

\newcommand{\id}{\ensuremath{\mathbbm 1}}

\newcommand{\BE}{\begin{equation}}
\newcommand{\EE}{\end{equation}}
\newcommand{\be}{\begin{equation}}
\newcommand{\ee}{\end{equation}}
\newcommand{\bea}{\begin{eqnarray}}
\newcommand{\eea}{\end{eqnarray}}
\newcommand{\bean}{\begin{eqnarray*}}
\newcommand{\eean}{\end{eqnarray*}}
\newcommand{\kommentar}[1]{}

\newcommand{\proj}[1]{\ketbra{#1}}
\newcommand{\tr}{{\rm Tr}}
\newcommand{\str}{{\rm str}}
\newcommand{\bc}{\begin{center}}
\newcommand{\ec}{\end{center}}
\newcommand{\proofend}{\hfill\fbox\\\medskip }

\newcommand{\rr}{\mathbbm{R}}

\newcommand{\cc}{\mathbbm{C}}

\newcommand{\newproof}{\noindent {{\em Proof. }}}

\newcommand{\jbox}[1]{
\begin{equation}
\text{#1}
\end{equation}}

\newtheorem{theorem}{Theorem}

\newtheorem{proposition}{Proposition}

\begin{document}

\title{Optimal entanglement witnesses for continuous-variable systems}

\author{P.\ Hyllus and J.\ Eisert}
\affiliation{1 QOLS, Blackett Laboratory, 
Imperial College London,
Prince Consort Road, London SW7 2BW, UK\\
2 Institute for Mathematical Sciences, Imperial College London,
Prince's Gardens, London SW7 2PE, UK}

\date{\today}

\begin{abstract}
This paper is concerned with all tests for 
continuous-variable entanglement that arise from 
linear combinations of second moments or variances 
of canonical coordinates, as they are commonly 
used in experiments to detect entanglement. All such tests
for bi-partite and multi-partite entanglement 
correspond to hyperplanes in the set of second moments. 
It is shown that all
optimal tests, those that are most robust against
imperfections with respect
to some figure of merit for a given state, 
can be constructed from solutions to
semi-definite optimization problems. 
Moreover, we show that for each such test, 
referred to as entanglement witness based
on second moments,
there is a one-to-one correspondence 
between the witness and a stronger product criterion,
which amounts to a  non-linear witness,
based on the same measurements.
This generalizes the known product criteria. The presented
tests are all applicable also to non-Gaussian states.
To provide a service to the community, we present the
documentation of two numerical routines, {\tt FullyWit} 
and {\tt MultiWit}, which have been 
made publicly available.
\end{abstract}


\maketitle

\section{Introduction}

The field of continuous-variable
quantum information has seen a very substantial 
progress in recent years. This has in part been 
made possible -- from the experimental
side -- by the availability of a number of sources of 
systems prepared in entangled states in the canonical 
coordinates. Notably, two-mode squeezed states of light
close to minimal uncertainty have been prepared 
\cite{Kimble1},
as well as bright entangled light beams 
\cite{Leuchs,LeuchsNeu}.
The collective spin states
of atomic ensembles have been brought into states that can
be well-described in terms of continuous-variable 
entanglement 
\cite{Polzik},
even allowing for a light-matter interface 
\cite{Interface1}.
First 
instances of multi-mode, multi-partite entangled states
have also been prepared already \cite{LoockExp}. 

In many of these set-ups, the starting point
of further exploitation of entanglement 
is to see whether the envisioned bi-partite or multi-partite 
entanglement can be found in the prepared states. This 
is often done by making use of criteria of entanglement based
on second moments, or uncertainties, of quantum states. 
Notably, in a two-mode set-up, the probably most well-known criterion
of this type is the following: if one finds that the state
$\rho$ of  a two-mode system
equipped with the canonical coordinates $\hat x_1$ and $\hat p_1$
of one mode and $\hat x_2$ and $\hat p_2$ for the second mode
fulfils
\begin{equation}\label{SimpleDuan}
	\langle (\hat u - \langle \hat u\rangle_\rho)^2\rangle_\rho
	+
	\langle (\hat v - \langle \hat v\rangle_\rho)^2\rangle_\rho
	< 1
\end{equation}
where $\hat u = (\hat x_1 + \hat x_2)/\sqrt{2}$ and $\hat v = (\hat p_1 - \hat p_2)/\sqrt{2}$,
then one can assert that the state must have been entangled 
\cite{Duan}. 
Such a criterion is tremendously helpful: firstly, it gives a clearcut 
test for deciding whether a state is entangled or in a subset where one 
cannot assert whether it was separable or entangled. Secondly, 
one only has to measure certain
fixed combinations of second moments of the original canonical 
coordinates. This is, yet, only one
specific test, detecting the entanglement
in some states, not detecting it in others. 
Similar tests have been proposed also to detect 
the entanglement of certain multi-party entangled states 
\cite{Peter,LoockExp}.

\begin{figure}[!h]
  \begin{center}
      \leavevmode
\resizebox{6 cm}{!}{\includegraphics{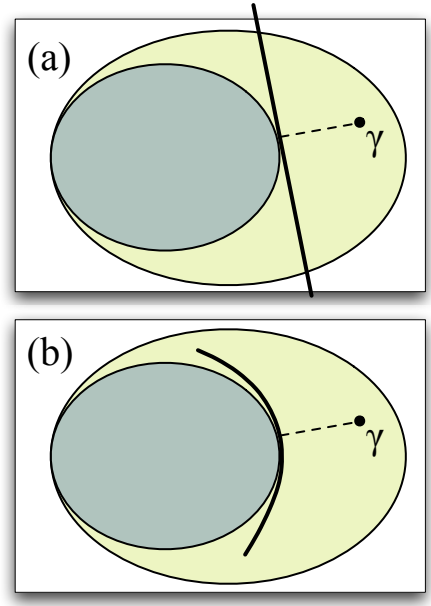}}

\end{center}
\caption{(a) Schematic representation of the optimal
entanglement witness based on second moments, with 
respect to a state with covariance matrix $\gamma$, as
a separating hyperplane from the convex
set of  second moments
consistent with separable Gaussian states (dark grey). (b)
Curved quadratic witness, which is curved towards the
above convex set.}
\label{coupling1}
\end{figure}

Commonly used tests for entanglement are entanglement witnesses 
\cite{witnesses}. An entanglement witness is an Hermitian 
operator $W$ such that $\tr[W\sigma_s]\ge 0$ holds for all 
separable states $\sigma_s$, while $\tr[W\rho]<0$ for some 
entangled state $\rho$ \cite{witnesses}. These tests are 
based on expectation values. In contrast, the tests mentioned above 
correspond to {\em entanglement witnesses based on second moments},
or equivalently, on `variances' or `uncertainties'. 
The covariance matrices collecting the second moments of the states that lead 
to a given fixed value on the left hand side of (\ref{SimpleDuan})
give rise to a hyperplane in the set of all second moments
embodying the correlations of the state, see Figure \ref{coupling1}. 
It is a separating hyperplane: all second moments that correspond to 
separable states -- as well as some corresponding to entangled states --
are on one side of the hyperplane. Hence the test
either confirms the presence of entanglement or returns an inconclusive
result. 

These criteria are different than, for example,
criteria directly based on the positivity of the
partial transpose (PPT) \cite{Simon}, requiring the full
knowlegde of all second moments. This PPT condition
then also constrains symplectic eigenvalues of covariance
matrices, and hence marginal and full 
purities \cite{Illu,Jaromir}.

It has been shown how the set of entanglement witnesses
in this sense can indeed be completely 
characterized \cite{GBook}. However, important open
questions remain: Given a state with some covariance matrix, 
can one find an optimal entanglement witness in the sense 
that it most robustly detects the state as being entangled?
This is important, since for experiments aiming at
the production of a specific entangled state, the answer would
deliver an optimal test detecting the entanglement, optimally
robust to noise.
Another question is whether this can 
be done for
all different separability classes versus multi-particle 
entanglement. This paper gives a positive answer to these questions. 
In particular, we show that all such optimal tests 
(and not only the test
whether a covariance matrix corresponds to a separable
Gaussian state \cite{Remark})
in the bi- and multi-partite setting arise 
from solutions to certain semi-definite problems 
\cite{SDP}.

To provide a service to the community, we present a publicly 
available software package, consisting of the functions {\tt FullyWit} 
and {\tt MultiWit}. Given the covariance matrix of a state, the first
one finds optimal witnesses detecting entanglement when a certain splitting 
of the parties sharing the state is held 
fixed, while the latter identifies 
witnesses detecting only genuine multi-partite entanglement. 
Further, the routines 
allow to restrict the types of measurements that one wants to 
perform. For example, for a two-mode squeezed 
state in the so-called standard
basis under no further constraints the test as in Eq.~(\ref{SimpleDuan}) 
would be delivered as output.
Hence all entanglement witnesses in the bi- and the multi-partite
setup are efficiently obtained. 

Further, we show that for each entanglement witness, there
is a one-to-one correspondence to a 
{\it curved quadratic witness}, thereby generalizing the 
known {\it product criteria} \cite{Mancini}, 
related to expressions of the type 
\begin{equation}\label{SimpleProduct}
	\langle (\hat u - \langle \hat u\rangle_\rho)^2\rangle_\rho
	\cdot
	\langle (\hat v - \langle \hat v\rangle_\rho)^2\rangle_\rho
	< \frac{1}{4}.
\end{equation}
These correspond to separating hyperplanes which are
curved towards the set of separable covariance matrices,
cf. Figure~\ref{coupling1}.
Also, the complete set of {\it generalized quadratic witnesses} is 
stated and discussed.

The paper is organized as follows: In Section \ref{sec:Prelims},
we recall the basic definitions regarding Gaussian states,
the classification of entangled states in a general setting, 
and semi-definite programs. In Section \ref{sec:witnesses} we state the definition 
of linear entanglement witnesses based on second moments. The 
semi-definite programs designed to find the optimal witnesses described 
above are presented in the Sections \ref{sec:SDP} and 
\ref{sec:multi}. We make some remarks on the form of the witnesses
when the state in question does not include correlations between
position and momentum variables and other issues in 
Section \ref{sec:remarks}. The theoretical part is concluded with
the characterization of the product witnesses in Section 
\ref{sec:curved}. The final Section \ref{sec:numerics}
contains several numerical examples where optimal witnesses
have been found for several states with the help of the functions
{\tt FullyWit} and {\tt MultiWit} based on the  results of the
Sections \ref{sec:SDP} and \ref{sec:multi}, respectively.

\section{Preliminaries}
\label{sec:Prelims}

\subsection{Definitions}

We  consider system consisting of $n$ modes,
associated with  {\it  canonical coordinates}
$\hat{r}=(\hat{x}_1,\hat{p}_1,\hat{x}_2,
\hat{p}_2,\ldots,\hat{x}_n,\hat{p}_n)$,
satisfying the canonical commutation relations, giving
rise to the familiar 
skew symmetric matrix
\begin{equation}
	\sigma=\bigoplus_{i=1}^n \left[
		\begin{array}{cc}
			0  & 1 \\
			-1 & 0
		\end{array}
	\right].
\end{equation}
These canonical coordinates will typically -- but not necessarily --
refer to amplitude and phase quadratures of 
a finite number of modes of the 
electromagnetical field of light.
In the following we will often consider the multi-partite 
situation: here, a subsystem 
$A$ embodies $n_A$ 
modes, system $B$ consists of $n_B$ modes, 
$C$ of $n_C$ modes, and so on, such that 
$n_A+n_B + ... =n$. In this paper, we 
will investigate entanglement and 
separability properties of such states on multi-partite 
systems. We will refer to a {\it split} as a coarse graining
of subsystems, i.e., a distribution of the $n$ physical
subsystems into groups that are considered the 
subsystems.

For our purposes the moments of the states will 
play the central role. The first moments are the 
displacements in phase space, 
$d_j=\langle \hat r_j\rangle_\rho$, $j=1,\ldots,n$.
The second moments, the variances, can be collected in
the  {\it covariance matrix} $\gamma$ of the state,
with entries
\begin{equation}
	\label{eq:coma}
	\gamma_{j,k}= 2 \Re \langle (\hat r_j-\langle \hat r_j\rangle_\rho) 
	(\hat r_k - \langle \hat r_k \rangle_\rho) 
	\rangle_\rho 
\end{equation}
$j,k=1,\ldots,n$. For a survey about these
preliminaries, see also Refs.\ 
\cite{CV,Review}. 
Any covariance matrix of a 
quantum state satisfies the 
Heisenberg uncertainty principle
\begin{equation}\label{Heisenberg}
	\gamma + i\sigma\geq 0.
\end{equation}
In turn, for every real symmetric matrix 
$\gamma\in \R^{2n\times 2n}$ satisfying (\ref{Heisenberg})
there exists a physical state with just these second moments
$\gamma$. Gaussian states -- those quantum states for which
the characteristic function is a Gaussian in phase space 
-- are uniquely characterized by 
their first and second moments
\cite{CV,Review}.  
Such Gaussian states 
play a central role in continuous-variable quantum
information, essentially since the Gaussian operations 
-- completely positive maps preserving the Gaussian
character -- 
are to a large extent
readily accessible \cite{Op,GiedkeOperations}.
However, the criteria we present can also detect 
the entanglement of non-Gaussian states.

It is a very useful fact that any matrix 
$M\in \rr^{2m\times 2m}$, $M > 0$, can be diagonalized as 
\begin{equation}
	S M S^T = D, 
\end{equation}
where $S\in Sp(2m, \rr)$
is not an orthogonal matrix, 
but one leaving the symplectic
form invariant, i.e., $S\sigma S^T =\sigma$. 
These canonical transformations are referred to as symplectic
transformations. The diagonal matrix $D$ can be taken to have 
the form $D= (s_1,s_1,\ldots,s_m,s_m)$ with $s_1,\ldots,s_m\geq 0$. 
These values are the {\it symplectic eigenvalues} 
of $M$ (different from the eigenvalues), which are also given by
the eigenvalues of the matrix 
$M^{1/2} (i \sigma) M^{1/2}$.
The symbol $\str$ denotes the symplectic trace of a matrix which is defined
as
\begin{equation}
	\str[M] = \sum_{j=1}^m s_m,
\end{equation}	 
counting each
symplectic eigenvalue only once. So it is essentially 
the trace of the matrix after symplectic diagonalization.

\subsection{Separability}
\label{sec:classification}

Quantum states of bi-partite systems may be classically
correlated or entangled. If they can be prepared by means
of local quantum operations and shared randomness alone,
a state is called {\it separable}. 
A state vector $\ket{\phi}$ in turn is entangled
if it cannot be written as as a tensor product 
$\ket{\phi}=\ket{a}\otimes\ket{b}$ of state vectors
$\ket{a}$, $\ket{b}$. A general mixed state of a bi-partite 
system is called separable \cite{Werner89} if it can be represented as a 
convex combination of products
\be
	\sum_i p_i \proj{\phi_i}\otimes\proj{\psi_i},
\ee
where $p_i\ge 0$ and $\sum_i p_i=1$. Otherwise, it is called
{\it entangled}.  

For quantum systems consisting of more than two
constituents, different kinds of classically correlated, i.e., of
separable states are conceivable. Depending on possible
preparation strategies, a state can be classified according
to a full hierarchy. At the lowest level of the hierarchy
are those states that contain no entanglement at all.
Such an $N$-partite (mixed) state $\rho$ is called 
{\it fully separable}, if it 
can be written as a convex combination of 
product states, so as
\be
 	\rho=\sum_k p_k \proj{\psi_i}^{(1)}
 	\otimes\proj{\phi_i}^{(2)}
 	\otimes\ldots\otimes\proj{\eta_i}^{(N)},
\ee
where $p_i\ge 0$ for all $i$ and $\sum_i p_i =1$. These
are states that can be prepared by means of local operations
with respect to all subsystems, together with shared
randomness. 

In a multi-partite system, yet, also other classes of 
separability are possible. To obtain a hierarchy, one
may consider $k$-partite splits, where each of the parts
is considered a subsystem in its own right, and refer to 
states that are fully separable with respect to such a $k$-partite
split as being $k$-{\it separable} \cite{Classification,Geza} (for
a short review, see also Ref.\ \cite{Gross}). 
Towards the end of this hierarchy are the 
$2$-separable or {\it bi-separable} states: they are
those states for which there exists a bi-partite 
split such
that the state is separable with respect to this split.
Needless to say, a state that is fully separable with
respect to one $k$-partite split might be entangled
when another $k$-partite split is considered.
Hence, all possible splits for all possible $k$ have
to be considered for a complete classification.
This classification is treated theoretically in 
Section \ref{sec:SDP} and practically by the 
routine {\tt FullyWit}.


According to this classification, 
a state is genuinely $N$-partite entangled if it is not
separable with respect to {\em any} split.
However, in general, there exist states which can be written
as a convex combination of certain 
$k$-separable states which
are not separable with respect to any split.
This motivates the alternative definition that an $N$-partite 
state is called genuinely $N$-partite entangled if it {\em cannot}
be written as a convex combination of some $k$-separable states 
\cite{Acin} for any $k \ge 2$.
For example, in a tri-partite system consisting of parts 
$A$, $B$, and $C$ only those states are then genuinely
three-partite entangled which cannot be written in the form
\bea\label{bis}
	\rho_\textrm{BS}&=&\lambda_1\sum_k 
	   p^{(A|BC)}_k\proj{\psi_k}^{(A)}\otimes\proj{\phi_k}^{(BC)}\\
	&+&\lambda_2\sum_j p^{(AB|C)}_j\proj{\psi_j}^{(AB)}
		\otimes\proj{\phi_j}^{(C)}\\
	&+&\lambda_3\sum_l p^{(AC|B)}_l\proj{\psi_l}^{(AC)}
	     \otimes\proj{\phi_l}^{(B)},
\eea
where $\lambda$, $p^{(A|BC)}$, $p^{(AB|C)}$, and $p^{(AC|B)}$
form probability distributions. 
Genuine $N$-partite entanglement
according to this definition is treated theoretically 
in Section \ref{sec:multi} 
and practically by the routine {\tt MultiWit}.

The definitions from above immediately carry over to continuous variable
systems with canonical coordinates. Here, they can be expressed in 
terms of covariance matrices.
Let $\gamma$ be the covariance matrix of a state on $n$ modes
with finite second moments, which is fully separable with 
respect to $n$ subsystems.
Then there exist covariance matrices $\gamma^{(i)}, i=1,\ldots,n$,
corresponding to the $n$ subsystems, such that \cite{CVbe}
\be
		\gamma\geq 
			\gamma^{(1)}\oplus\ldots\oplus\gamma^{(n)}.
		\label{separability}
\ee
Conversely, if this holds, then Gaussian states with the covariance 
matrix $\gamma$ are separable. Hence if this criterion is
violated, then the corresponing state is entangled, 
irrespective of whether it is Gaussian or not. If it is not violated, 
then a Gaussian state is separable while a 
non-Gaussian state might 
be entangled. Keeping this in mind we will call all covariance 
matrices fulfilling Eq.~(\ref{separability}) fully separable
to keep our notation simple.

Note that the problem of testing whether 
(\ref{separability}) can be satisfied is a semi-definite problem
in its own right \cite{Remark}, see the subsequent subsection. 
It is a feasibility problem, so the question
is whether or not such matrices $\gamma^{(1)},\ldots,\gamma^{(n)}$ 
can be found satisfying in turn the semi-definite constraints
$\gamma^{(j)}+ i \sigma\geq 0$ for all $j=1,\ldots,n$.

This statement can be generalized to bi-separable states
in the sense of a convex combination of pure bi-separable
states. 
Let $\gamma_\textrm{BS}$ be the covariance matrix 
of a bi-separable $n$-partite
state with finite second moments. Then there exist 
partitions $\pi$ of the
$n$ modes into two subsystems consisting of $m<n$ modes and $n-m$ modes, 
covariance matrices $\gamma_{\pi(k)}$ which are block diagonal 
with respect to the partition $\pi(k)$, 
and a probability distribution $\lambda$ so that
\begin{equation}
		\gamma_\textrm{BS}-\sum_k\lambda_k\gamma_{\pi(k)}\ge 0.
		\label{biseparability}
\end{equation}
Conversely, if this holds, then Gaussian states with the covariance 
matrix $\gamma_\textrm{BS}$ are bi-separable. 
The same comment on non-Gaussian states that we made 
after the criterion for full separability applies here
which we have to keep in mind when we call all covariance 
matrices fulfilling Eq.~(\ref{biseparability}) bi-separable.
Note further that these states include those with are separable 
for a split with $k>2$, hence all states which are not in this set
are genuinely $N$-partite entangled according to the
second definition.

\subsection{Semi-definite problems}

Semi-definite programs (SDP) \cite{SDP} will play a prominent
role in the subsequent argument. Readers familiar with the
subject may wish to jump to the next section.
Semi-definite programs
are convex optimization problems \cite{VBCO}
of a specific form: one minimizes a linear function, subject to a 
semi-definite constraint. Many problems in quantum information 
science can be cast into this form \cite{Many}, essentially originating from the
fact that semi-definite constraints appear in conditions to quantum
states, as well as to quantum operations via the duality between
positive operators and completely positive maps. Also, even
global optimization problems can be relaxed to semi-definite
form, with several applications to quantum information 
problems \cite{Relax1,Relax2}.

More specifically, a semi-definite program (SDP) is an optimization
problem of the following kind:
\bea
	\text{minimize}_x && c^T x \\
	\text{subject\ to} && F(x)=F_0+\sum_{i=1}^t F_i x_i \ge 0,\nonumber
\eea
where the minimization is performed with respect to a real vector
$x$ of length $t$. The problem is specified by the vector $c\in\R^t$
and the Hermitian matrices $F_i\in \C^{s\times s}, i=1,\ldots,t$.
This is the form that is usually referred to as being the
primal problem. It is desirable to 
formulate a given problem as an SDP, not the least
because these can be solved efficiently, for instance by 
using interior point  methods \cite{SDP}. 

Via Lagrange-duality the Lagrange-dual of the above 
problem can be formulated. Dual problems to SDPs are
again SDPs, where essentially the roles of objective
variables and constraints are interchanged. 
The so-called dual problem can be formulated as follows
\bea
	\text{maximize}_Z && -\tr[F_0 Z] \\
	\text{subject\ to} && Z \ge 0, \nonumber \\
			&& \tr[F_i Z]=c_i.\nonumber
\eea
The objective value of 
every solution of the dual provides a lower
bound to the value of any solution to the primal
problem and vice versa. 
This is referred to as {\it weak duality}:
For feasible $x$ and $Z$, i.e. $x$ and $Z$ 
fulfilling the respective constraints,
\be
  c^T x+\tr[F_0 Z]=\tr[F(x)Z]\ge 0
  \label{weakD}
\ee
holds, where the inequality is due to the fact
that $F(x)\ge 0$ and $Z\ge 0$. 
If either the primal or the dual problem (or both)
are strictly feasible, meaning that there
exists a feasible vector $x$ such that
$F(x)>0$ or there exists a feasible $Z>0$, 
then there exist $x^*$ and $Z^*$ such that 
\begin{equation}
	c^T x^*=-\tr[F_0 Z^*]. 
\end{equation}
This is referred to as
{\it strong duality}.

An important class of problems are the
{\em feasibility} problems. Here, $c=0$, 
so that the primal problem amounts to 
checking whether 
there exists any feasible $x$ fulfilling the primal
constraints. In this case, $\tr[F_0 Z]\ge 0$
has to hold for all feasible $Z$.
Hence, if there is a feasible $Z$ with $\tr[F_0 Z] < 0$,
then the primal problem cannot be feasible.

\section{Linear entanglement witnesses based on second moments}
\label{sec:witnesses}

\subsection{Concept of entanglement witnesses based on second moments}

We now turn to {\it entanglement witnesses based on second 
moments}. They are tests for entanglement based on linear
combinations of second moments. Each test corresponds to a
hyperplane in the set of second moments. In turn, these hyperplanes
encode the physical set-up, the type of measurement that is being 
physically performed. Such 
hyperplanes are defined by 
a real symmetric positive semi-definite (PSD) 
matrix $Z$ and a number $c\in \rr$,
via the Hilbert Schmidt scalar product. 
The hyperplane consists of all $\gamma$ such that 
\begin{equation}
	\tr[Z \gamma]=c.
\end{equation}
So an  {\it entanglement witnesses based on second 
moments} can be characterized by a real matrix $Z\geq 0$ 
satisfying \cite{GBook}
\bea
	\label{eq:FWit1}
	\text{(i)} && \tr[Z\gamma_s]\ge 1\quad \text{for\ all\ (fully)\ separable\ $\gamma_s$,}\\
	\label{eq:FWit2}
	\text{(ii)} && \tr[Z\gamma] < 1\quad \text{for\ some\ entangled\ $\gamma$.}
\eea
As mentioned in the introduction, this notion is very similar
to the one for entanglement witnesses based on expectation values: 
note that here, however, the witness refers to {\it second moments} 
of quantum states. If condition (ii) is fulfilled, then
the quantum state giving rise to $\gamma$ is entangled, 
irrespective of whether it is Gaussian or not, as discussed in
Sec.~\ref{sec:classification}.

These hyperplanes are hence separating entangled
covariance matrices from the set of fully separable covariance 
matrices. That this characterization is possible is
due to the fact that 
%
the set of fully separable covariance matrices is convex and closed. 
Its boundary is given by the matrices $\oplus_{k=1}^{n}\gamma_k$ 
fulfilling the condition $\oplus_{k=1}^{n}\gamma_k\ge i\sigma$.
It is clearly convex, because the semi-definite constraint
defining fully separable covariance matrices is
preserved under convex combination: if 
$\oplus_{k=1}^{n}\gamma_k\ge i\sigma$ and 
$\oplus_{k=1}^{n}\eta_k\ge i\sigma$ then 
\begin{equation}
	\alpha \oplus_{k=1}^{n}\gamma_k
	+(1-\alpha) \oplus_{k=1}^{n}\eta_k \ge i\sigma
\end{equation}
for $\alpha\in [0,1]$.
Further, the set is closed.
Firstly, the subset of covariance matrices of the form
$\oplus_{k=1}^{n}\gamma_k$ is closed itself since 
its complement is open: if a matrix $\gamma$ has 
nonvanishing off-diagonal elements, then in its neighborhood there
will be only matrices with nonvanishing off-diagonal 
elements. Secondly, 
the constraint 
\begin{equation}
	\oplus_{k=1}^{n}\gamma_k\ge i\sigma
\end{equation}
defines a closed convex cone which is a subset
of the space of matrices $\oplus_{k=1}^{n}\gamma_k$.
Because the set is convex and closed, there
exist hyperplanes separating a covariance
matrix $\gamma$ which is not separable from
the set of separable covariance 
matrices \cite{VBCO}.

\subsection{Familiar examples}

Any such matrix $Z$ encodes the measurement pattern
performed for a certain test, so the linear combination of 
second moments that is required to assert that a state was
entangled. For example, the matrix $Z$ for the familiar
test of Eq.\ (\ref{SimpleDuan}) from Ref.\ \cite{Duan} can 
be written as
\begin{equation}
	Z=\frac{1}{4}\left[
	\begin{array}{cccc}
	1 & 0 & 1 & 0 \\
	   0 & 1 & 0 & -1\\
	   1 & 0 & 1 & 0\\
	   0 & -1 & 0 & 1\\
	\end{array}
		\right].
\end{equation}
Using the definition (\ref{eq:coma}) it follows that 
\be 
	\tr[Z\gamma]=
	\langle (\hat u - \langle \hat u\rangle_\rho)^2\rangle_\rho
	+
	\langle (\hat v - \langle \hat v\rangle_\rho)^2\rangle_\rho,
\ee
such that inequality~(\ref{SimpleDuan}) is equivalent with
\begin{equation}
	\tr[Z\gamma] <1,
\end{equation}
from which one can conclude that the state must have been
entangled. In turn, the test gives rise to a hyperplane
in the space of second moments separating a subset of
all entangled states from the separable states.
The general criterion of Ref.\  \cite{Duan}, dependent on a parameter 
$a\in \rr\backslash\{0\}$, reads as
\be
	\langle (\hat u_a - \langle \hat u_a\rangle_\rho)^2\rangle_\rho
	+
	\langle (\hat v_a - \langle \hat v_a\rangle_\rho)^2\rangle_\rho
	\ge a^2+\frac{1}{a^2}
\ee
for all separable $\rho$, where 
\be
	\hat{u}_a=|a|\hat{x}_1+\frac{1}{a}\hat{x}_2,\quad \hat{p}_a=|a|\hat{p}_1-\frac{1}{a}\hat{p}_2.
	\label{eq:DuanGen}
\ee
This test corresponds to a particular
hyperplane  for each $a$,
corresponding to the entanglement witness
\begin{equation}
	\label{eq:DuanWit}
	Z=\frac{1}{2(a^2+\frac{1}{a^2})}\left[
	\begin{array}{cccc}
	a^2 & 0 & \frac{|a|}{a} & 0 \\
	   0 & a^2 & 0 & -\frac{|a|}{a}\\
	   \frac{|a|}{a} & 0 & \frac{1}{a^2} & 0\\
	   0 & -\frac{|a|}{a} & 0 & \frac{1}{a^2}\\
	\end{array}
		\right],
\end{equation}
so that 
\bea
	&& \tr[Z_a\gamma_s]\ge 1\ \text{for\ all\ separable\ }\gamma_s,\\
	&& \tr[Z_a\gamma]<1\ \text{for\ some\ entangled\ }\gamma.
\eea
In Ref.~\cite{Simon,Duan} it was further shown that the covariance 
matrix can always be brought into the standard form of a direct sum in 
position and momentum variables, using local symplectic transformations.
The test~(\ref{eq:DuanGen}) is necessary and sufficient for
covariance matrices in this standard form.
In other words, if the covariance matrix of a state is known,
and we allow for further local operations based on that knowledge,
then the set of witnesses $Z_a$ completely 
characterizes the set of 
separable Gaussian states.
%

Another example of criteria that correspond to witnesses
based on second moment are the tests for multi-partite 
entanglement of Ref.~\cite{Peter}.
In turn, the set of witnesses of second moments can be fully 
characterized \cite{GBook}, as sketched in the Appendix.

\subsection{Detecting entanglement in non-Gaussian states}

There is finally a comment in order concerning {\it non-Gaussian states}:
As mentioned before, all such tests also detect non-Gaussian states
as being entangled. This is due to the fact that if a Gaussian state is entangled,
then every non-Gaussian state with the same second moments is 
necessarily entangled. Hence, one does not have to assume a priori
that the state 
in question
is exactly Gaussian. This knowledge is typically not available anyway, 
without referring to full quantum state tomography. 

One has to be aware, however, that in 
such infinite-dimensional quantum systems, the entangled
states are trace-norm dense in state space \cite{Infinity}. That is,
in any neighborhood of a separable state an
entangled state can be found, and the volume of the set
of separable states has measure zero. 
This situation remains unaltered
if one introduces a constraint to the mean energy 
of the system \cite{Infinity}.

Nevertheless, it still
makes sense to test for entanglement
in this setting: one only has to state the result in the 
form that if a state is detected as being entangled then
a trace-norm ball centered at it is not 
consistent with any separable state. In other words,
one can still meaningfully say that a state is `far 
from being separable'.

\section{Task of finding witnesses as semi-definite problem}
\label{sec:SDP}


The common situation that one encounters is the
following: one knows what kind of entanglement one would
like to see in a certain state, prepared in some setup. Also, one
typically 
has an idea about how the covariance matrix $\gamma$ of 
the prepared state roughly looks like or at least about how 
it is desired to look like. 
The question is: what
measurements have to be performed in order to most easily
detect the entanglement? More specifically, in case of 
bi-partite entanglement, the task is the following: 

We have two parts consisting of $n_A$ and $n_B$ modes, 
respectively.
For a given $\gamma$ (the covariance matrix that we suspect that we 
have) corresponding to an entangled Gaussian state
we would like to find the test $Z$ with the property that $Z$ 
detects $\gamma$ as corresponding to an entangled state, 
such that 
\begin{equation}
	w=\tr[Z\gamma]<1
\end{equation}
takes its minimal value. 
This is the test which `most distinctly' detects $\gamma$ as
originating from an entangled state, in a way that is most robust against
detection imperfections. Geometrically, we aim at finding the hyperplane with
the greatest distance from $\gamma$. Obviously, not only $\gamma$ is 
detected as coming from an entangled state by this test, but the test
is optimized for this specific guess.

\subsection{The primal problem}

It turns out that the previous 
problem is related to the following 
optimization problem. For a separable $\gamma$,
we have that there exist covariance matrices
$\gamma_A$ and $\gamma_B$, satisfying the Heisenberg
uncertainty relation
$\gamma_A\oplus\gamma_B+ i \sigma\geq 0$. For 
covariance matrices $\gamma$ corresponding to
entangled states, we may 
write the primal problem in the following form:
\begin{eqnarray}
	 \textrm{minimize}_{\gamma_A,\gamma_B,x_e} && ( -x_e),\\
	 \textrm{subject to} 
	&& \gamma-\gamma_A\oplus\gamma_B\ge 0,	\nonumber\\
	&& \gamma_A\oplus\gamma_B+(1+x_e)i\sigma\ge 0.\nonumber
\eea
If there is an optimal 
solution with $x_e\ge 0$, then $\gamma$
is separable, because $\gamma_A\oplus\gamma_B$ 
fulfils an even stricter form of the Heisenberg uncertainty
relations. If $x_e<0$, then $\gamma$ is entangled,
since $\gamma_A\oplus\gamma_B$ 
can now violate the uncertainty relations. This is actually 
just the $p$-measure 
from Refs.\ \cite{GiedkeOperations}, up to $p=1/(1+x_e)$.
For Gaussian states, $-\log_2(p)$ is a lower bound for 
the {\it logarithmic negativity}, defined as 
\begin{equation}
	E_N(\rho)= \log_2\|\rho^\Gamma\|_1, 
\end{equation}
where
$\rho^\Gamma$ denotes the partial transpose of 
$\rho$ and $\|.\|_1$ is the trace norm. It is moreover
identical to the logarithmic negativity
for $1\times n$-mode systems \cite{GiedkeOperations}.
The negativity \cite{LogNeg}
is a measure of entanglement, 
and indeed a monotone under local operations and classical
communication \cite{PhD,Vidal,Plenio}.

\subsection{The dual problem}

The dual problem of the above problem can easily be found. 
The key point in what follows 
is that from the dual of the above
semi-definite problem allowing for a pre-factor in the Heisenberg 
uncertainty one can extract the required optimal tests.
The dual problem can be cast into the form
\bea
	\label{eq:fullydual1}
	\text{maximize}_X && -\tr[(\gamma\oplus i\sigma) X]\label{eq:witcon1}  \\
	\text{subject\ to} && X \ge 0, \nonumber \\
		&& \tr[(0\oplus i\sigma) X]=-1,\nonumber  \\
		&& \tr[(-F_{j,k}\oplus F_{j,k} ) X]=0, \nonumber\\
		&& 	\hspace{2cm} j,k=1,\ldots,n_A,   \nonumber\\
	  && \tr[(-F_{j,k}\oplus F_{j,k} ) X]=0, \nonumber\\  
	  && \hspace{2cm} j,k=n_A+1,\ldots,n,  \nonumber 
\eea
where the maximization is performed over
{\em Hermitian} matrices $X\in\cc^{4n\times 4n}$.
This matrix corresponds to a partitioning of 
degrees of freedom labeled $1,\ldots,n_A$ of system $A$ first, 
then $n_A+1,\ldots,n$ labeling the modes of system $B$, and
then again the same ordering.
$F_{j,k}$, $j,k=1,\ldots,n$, form a set of
real symmetric matrices all entries of which are
zero, except
\begin{equation}
	(F_{j,k})_{j,k}=(F_{j,k})_{k,j}=1. 
	\label{eq:sbasis}
\end{equation}
These matrices $F_{j,k}$ form a basis of all real symmetric 
$n\times n$-matrices.
The form of Eq.~(\ref{eq:fullydual1}) of the dual problem becomes 
manifest when expressing the
primal problem in terms of this operator basis, and
writing the two semi-definite constraints of the primal
problem in form of a direct sum as one constraint.

Due to the block diagonal structure of 
$(\gamma\oplus i\sigma)$ and all constraints, 
we can without loss of generality assume that 
\begin{equation}
	X=X_1\oplus X_2.
\end{equation}
The first constraint is then equivalent to $X_{1,2}\ge 0$. 
The latter constraints in the dual problem lead to 
$ \tr[ F_{j,k}X_1] =  \tr[F_{j,k}  X_2]$,
which restricts the real symmetric
single system blocks of system $A$ and $B$ in
$X_1$ and $X_2$ to be equal. 
This generalizes directly to the case of $N$ subsystems:
then the $N$ real
symmetric single system blocks of $X_1$ have to be 
equal to the $N$ real symmetric single party blocks 
of $X_2$.
The matrix $X_2$ is further restricted by the
condition
\begin{equation}
	\tr[(0\oplus i\sigma) X]=\tr[i\sigma X_2]=-1.
\end{equation}	
Finally, the dual objective function is
\begin{equation}
  \tr[(\gamma\oplus i \sigma)X]
  =\tr[\gamma X_1]+\tr[i\sigma X_2]
  =\tr[\gamma X_1]-1.
\end{equation}
Since $\gamma$ is real and symmetric, we have further that
$\tr[\gamma X_1]=\tr[\gamma X_1^\textrm{re}]$, where $X_1^\textrm{re}$
is the real part of $X_1$.

To summarize, the dual problem can be
formulated as
\bea
	\label{eq:dual1}
	\text{minimize}_{X_1,X_2} && \tr[\gamma X_1^\text{re}]-1 ,\\
	\nonumber
	\text{subject\ to} && X_{1}^\text{bd,re}=X_{2}^\text{bd,re}\\
	\nonumber
	&& X_1 \ge 0,\ X_2\ge 0,\\
	&& \tr[i\sigma X_2]=-1.\nonumber
\eea
In this formulation, no basis is used 
explicitly. $X^{\text{bd}}_{1,2}$
refers to the block diagonal matrices obtained
from $X^{\text{bd}}_{1,2}$ through pinching to the blocks
of system $A$ and $B$ (i.e., a projection onto the form of a 
direct sum).  
Now we are in the position to 
formulate the connection between
the separability problem and witnesses
based on second moments:
\begin{proposition}[Optimal witnesses in bi-partite systems]
For every feasible solution $X$ to
the dual program formulated above, the
matrix $X_1^\textrm{re}$ fulfils the witness 
condition (\ref{eq:FWit1}).
If $\gamma$ is entangled, then 
\begin{equation}
	\tr[\gamma X_1^\textrm{re}]<1, 
\end{equation}
so that $X_1^\textrm{re}$ also fulfils 
condition (\ref{eq:FWit2}).
Further, $\tr[\gamma X_1^\textrm{re}]$
is the minimal value of 
$\tr[\gamma Z]$ for any 
witness $Z$.
\end{proposition}

\newproof
From weak Lagrange duality (\ref{weakD}) it 
follows that
\begin{equation}
	c^T x+\tr[(\gamma\oplus i\sigma) X] \ge  0,
\end{equation}
where $c$ is the vector specifying the objective function of
the primal problem, 
being equivalent with
\begin{equation}	
	\tr[\gamma X_1^\textrm{re}]\ge 1+x_e.
	\label{eq:weakD2}
\end{equation}
In this case, there is always a strictly 
feasible $X$: just take 
$X=\Eins\oplus(\Eins-i\sigma/(2n_A+2n_B))$.
Hence there exist feasible $X$ and $x$
such that equality is obtained in Eq.~(\ref{eq:weakD2}). 
We have seen before that $x_e\ge 0$ 
for all separable states. Hence the 
condition~(\ref{eq:FWit1}) is fulfilled.
On the other hand, if $\gamma$ is an
entangled covariance matrix, then $x_e<0$,
hence also the condition~(\ref{eq:FWit2})
if fulfilled.
Since the equality holds in Eq.~(\ref{eq:weakD2}),
$\tr[\gamma X_1^\textrm{re}]$ reaches the
minimal value.
\proofend

An analogous proposition holds for witnesses
detecting full separability in a system of $N$ 
subsystems. Hence all classes of $k$-separability
can be tested with this criterion.


\section{Detecting genuine multi-partite entangled states}
\label{sec:multi}

The previous section was devoted to tests of full separability.
Here, we formulate the problem for excluding bi-separability
for systems of $N$ subsystems, each consisting of
 $n_i$ modes, $i=1,...,N$. 
 As before, the total 
 number of modes is $n=\sum_i n_i$.
The condition bi-separable states have to fulfil is
inequality~(\ref{biseparability}).

\subsection{Primal problem}
We write the primal problem in the following form: 
\bea	
 	\textrm{minimize}_{\{\gamma_{\pi(k)}\},x_e} && -x_e,\\
	 \textrm{subject\ to}
	&& \gamma-\sum_k\gamma_{\pi(k)}\ge 0\nonumber\\
	&& \gamma_{\pi(k)}+\lambda_{k}i\sigma \ge 0\ \textrm{for\ all\ } k,	
		\nonumber\\
	&& \sum_k  \lambda_{k}=1+x_e,\nonumber\\
	&& \lambda_{k}\ge 0\ \textrm{for\ all\ } k.\nonumber
\eea
Here $\pi(k)$ are all $K=2^{N-1}-1$ possible bi-partite partitions 
of the $N$ systems. The matrices $\gamma_{\pi(k)}$
are block diagonal with respect to the partition $\pi(k)$.
If the solution $x_e\ge 0$, then $\gamma$
is bi-separable, because the matrices $\sum_k\gamma_{\pi(k)}$ 
fulfil an even stricter form of the Heisenberg uncertainty
relations. If the solution $x_e<0$, then $\gamma$ is
genuinely multi-partite entangled, since 
$\sum_k\gamma_{\pi(k)}$ can now violate the uncertainty
relations.

With the basis introduced in the previous section,
the problem can be formulated as 
\bea	
  \textrm{minimize}_{\{x^{\pi(k)}_{i,j}\},x_e} && -x_e,\\
   \textrm{subject to}&&
 \gamma+\sum_{k,i,j}^\text{bd,re,$\pi(k)$}(-F_{i,j})
 x^{\pi(k)}_{i,j}\ge 0,\nonumber\\
	&& \sum_{i,j}^\text{bd,re,$\pi(k)$} F_{i,j}
	x^{\pi(k)}_{i,j}+\lambda_{k} i\sigma\ge 0\ 
	\textrm{for\ all\ } k,\nonumber\\
	&& \sum_k\lambda_k-x_e-1\ge 0, 	\nonumber\\
	&& -(\sum_k\lambda_k)+x_e+1\ge 0, 	\nonumber\\
	&& \lambda_{k}\ge 0\ \textrm{for\ all\ } k,\nonumber
\eea
where the index `bd,re,$\pi(k)$' 
refers to `block-diagonal and real with respect to
the partition $\pi(k)$'.
The set of constraints can again be cast into the form of a 
single constraint in a direct sum form. This helps to
identify the respective terms in the dual problem.

\subsection{Dual problem}

We can again assume the Hermitian matrix 
$X\in \cc^{(2n[K+1]+2+K)\times (2n[K+1]+2+K) }$ to
be block diagonal. The dual problem is then given by
\bea
	\label{eq:Mdual1}
  \textrm{maximize}_X && -\tr[(\gamma\oplus 0_{2nK}
  \oplus (-\id_1) \oplus \id_1 \oplus 0_K) X],\\
   \textrm{subject\ to}
  && X_1^\textrm{re,bd,$\pi(k)$}=X_{k+1}^\textrm{re,bd,$\pi(k)$}\ \textrm{for\ all\ }k,\nonumber\\
  && \tr[i\sigma X_{k+1}]+X_{K+2}-X_{K+3}+X_{K+3+k}=0
  \nonumber \\ 
  && \;\;\;\;\;\;\; \textrm{for\ all\ }k,\nonumber\\
  && X_{K+2}-X_{K+3}=1,\nonumber
\eea
where $\Eins_d$ and $0_d$ are the $d$-dimensional
identity operator and $0$ operator, respectively.
With the last constraint, the objective function reduces to
\begin{eqnarray}
	&&\tr[(\gamma\oplus 0_{2nK}
  \oplus (-\id_1) \oplus \id_1 \oplus 0_K) X]  
	\nonumber\\
	&&\;\;\;\; =\tr[\gamma X_1^\textrm{re}] - (X_{K+2}-X_{K+3})\nonumber\\
	&&\;\;\;\; = \tr	[\gamma X_1^\textrm{re}]-1.
\end{eqnarray}	
Now we can formulate the connection between
the bi-separability problem and witnesses:

\begin{proposition}[Witnesses for multi-partite entanglement]
For every feasible solution $X$ to
the dual program formulated above, the
matrix $X_1^\textrm{re}$ satisfies the witness 
condition 
\begin{equation}
	\tr[X_1^\textrm{re}\gamma_\textrm{BS}]\ge 1
\end{equation}
for all bi-separable $\gamma_\textrm{BS}$.
If $\gamma$ is genuinely multi-partite entangled, then 
$\tr[\gamma X_1^\textrm{re}]<1$ so that
$X_1^\textrm{re}$ also satisfies
condition (\ref{eq:FWit2}).
Further, $\tr[\gamma X_1^\textrm{re}]$
is the minimal value of 
$\tr[\gamma Z]$ for any 
witness $Z$ detecting
only genuinely multi-party entangled states.
\end{proposition}

\newproof
From weak Lagrange duality~(\ref{weakD}) it  follows that 
\bea
	c^T x+\tr[(\gamma\oplus 0_{2nK}
  \oplus (-\id_1) \oplus \id_1 \oplus 0_K) X]& \ge & 0,
\eea
which is equivalent with
 \bea 
	\tr[\gamma X_1^\textrm{re}]\ge 1+x_e.
	\label{eq:weakD3}
\eea
In this case, there is always a strictly 
feasible $X$: just take 
\begin{eqnarray}
X_1&=&\Eins, \;\; X_{k+1}=\Eins-\frac{1}{n}i\sigma,\\
X_{K+2}&=&\frac{3}{2},\;\; X_{K+3}=\frac{1}{2}, \\
X_{K+3+k}&=&1
\end{eqnarray}
 for all $k$. Here, $n$ is the total number of modes as before.
Hence there exist feasible $X$ and $x$
such that equality is obtained in the 
last equation. We have seen before that $x_e\ge 0$ 
for bi-separable states. Hence 
$X_1^\textrm{re}$ fulfils the condition~(\ref{eq:FWit1}).
Further, if $\gamma$ is genuinely multi-partite entangled,
then $x_e<0$. Hence $X_1^\textrm{re}$ also respects 
the condition~(\ref{eq:FWit2}). Finally,
since the equality holds in Eq.~(\ref{eq:weakD3}),
$\tr[\gamma X_1^\textrm{re}]$ reaches the
minimal value.
\proofend


\section{Remarks}
\label{sec:remarks}

\subsection{Direct sum form of tests for block diagonal covariance matrices}
\label{sec:bd_xp}

Often, the covariance matrices of generated states exhibit a direct
sum form with respect to position and momentum variables. 
In the simplest two-mode form, this corresponds to a 
covariance matrix of the form
\begin{equation}
	\label{eq:easystates}
	\gamma=
	\left[
	\begin{array}{cccc}
	\xi_1 & 0 & \xi_5 & 0 \\
	0 & \xi_2 & 0 & \xi_6\\
	\xi_5 & 0 & \xi_3 & 0\\
	0 & \xi_6 & 0 & \xi_4
	\end{array}
	\right],
\end{equation}
with some $\xi_1,\ldots,\xi_6\in \rr$. Not only that every two-mode
covariance matrix can be brought into such a form by means
of appropriate local symplectic transformations, but many 
of the second moments of typical generated states exhibit 
approximately this form anyway. Two-mode squeezed states or
noisy variants have covariance matrices
of this form, needless to say. Then, the question is: can the test also
be taken to be of this form, without losing optimality? The well-known
test in (\ref{SimpleDuan}) from Ref.\ \cite{Duan}, 
to give an example, is of such a form.
This question can be positively answered. 

\begin{proposition}[Direct sum in position and momentum]
Let $\gamma$ be a covariance matrix of the form of a direct
sum of matrices corresponding to position and momentum
coordinates (as in Eq.\ (\ref{eq:easystates}) for two modes).
If $Z$ is the optimal solution of the dual problem for $\gamma$, 
then the pinching
\begin{equation}
	Z'=P_p Z P_p + P_x Z P_x,
\end{equation}
where 
\begin{eqnarray}
	P_p&=&\text{diag}(0,1,\ldots,0,1),\label{Px}\\
	P_x&= & \text{diag}(1,0,\ldots,1,0)\label{Pp}
\end{eqnarray}	
is also a feasible solution with the same objective 
value $\tr[\gamma Z]$. 
\end{proposition}

\newproof
The first steps of the proof are straightforward:
Since $\gamma$ is of the form of a direct sum of 
matrices corresponding to the position
and momentum coordinates, 
$\gamma=P_x \gamma P_x + P_p \gamma P_p$ holds, 
and hence 
$\tr[Z\gamma]=\tr[Z'\gamma]$.
Now it has to be shown that $Z'$ is a witness.
First, it fulfils $Z'\ge 0$, as every principal submatrix
of a positive matrix is positive.
Secondly, $\tr[Z'\gamma_s]\ge 1$ 
has to hold for all separable $\gamma_s$. 
This is equivalent to $\tr[Z'\gamma_s]=\tr[Z\gamma_s']\ge 1$,
where 
\begin{equation}
	\gamma_s'=P_x \gamma_s P_x + P_p \gamma_s P_p.
\end{equation}
Hence if $\gamma_s'$ is a separable 
covariance matrix then $Z'$ is a witness.
The covariance matrix $\gamma_s$ fulfils 
$\gamma_s-\gamma_A\oplus\gamma_B\ge 0$ for some
covariance matrices 
$\gamma_A,\gamma_B$. Then, clearly
$\gamma_s'-\gamma_A'\oplus\gamma_B'\ge 0$ 
holds. It remains to show that $\gamma_A'\oplus\gamma_B'+i\sigma\ge 0$
if $\gamma_A\oplus\gamma_B+i\sigma\ge 0$.
Due to the block-diagonal structure, it
suffices to show that for any covariance $\eta$
satisfying $\eta \geq i \sigma$, 
also $\eta' \geq i \sigma$ holds. It is in this context
convenient to order coordinates as
$(\hat{x}_1,\ldots,\hat{x_n},\hat{p}_1,\ldots,\hat{p}_n )$.
Then, $\eta'$ is obtained from  $\eta$ as a result
of a pinching. In the following, we use standard notation from
Ref.~\cite{CV}. The idea of the proof is that we use
appropriate symplectic transformations that commute
with both $P_x$ and $P_p$ to transform
$P_x \eta P_x+P_p\eta P_p +i\sigma \mapsto 
P_x M P_x+P_p M P_p +i\sigma$, 
such that the problem is reduced to a
single mode problem. First, the position part
of $\eta$ can be brought to diagonal form 
by the congruence $\gamma\mapsto (O\oplus O)
\gamma (O\oplus O)^T$,
where $O\in O(n)$. Therefore, $[O,P_x]=[O,P_p]=0$. 
Then, with single mode squeezings,
\begin{equation}
	S= \text{diag}(d_1,\ldots,d_n,1/d_1,\ldots, 1/d_n), 
\end{equation}
$d_i\in\rr\backslash
\{0\}$ for $i=1,...,n$, the position part can be 
made proportional to the identity. 
Again, $[S, P_x]=[S,P_p]=0$.
Finally, the momentum part can be made diagonal, using
an appropriate $V\oplus V$, $V\in  O(n)$,
again of the form of $O$ from the first step, leaving the
upper block invariant. Hence, 
\begin{equation}
	M= (V\oplus V) S (O\oplus O) \eta (O\oplus O)^T S^T
	   (V\oplus V)^T
\end{equation}
is diagonal in both the position
and the momentum part. 

We can apply a pinching such that also the 
off-diagonal part of $M$ is diagonal, leaving $\sigma$
invariant. Since $M+i\sigma\ge 0$ holds,
then $M'+i\sigma \ge 0$ holds as well, where
$M'$ is the pinched form of $M$.
The covariance matrix is now 
a direct sum of single modes. 
But then, the validity of the 
statement becomes
obvious: if 
\begin{equation}
	\left[
	\begin{array}{cc}
	a & c\\
	c & b 
	\end{array}
	\right] + 
	i \sigma \geq 0,
\end{equation}
then always also 
\begin{equation}
	\left[
	\begin{array}{cc}
	a & 0\\
	0 & b 
	\end{array}
	\right] +
	i \sigma \geq 0
\end{equation}
holds true.
Hence, finally we arrive at
 $\gamma_A'\oplus \gamma_B'\geq i \sigma$.
\proofend

Hence, it does not restrict generality for covariance matrices
$\gamma$ in form of a direct sum of position and momentum contributions
to take a test with the same form. In any case, even if $\gamma$ does not
have this form, one may look at such tests, which we will later see
again in the context of product criteria.

\subsection{Incorporating practical measurement constraints}

Often, some combinations of second moments are
more accessible via experiment than others. One is 
typically well-advised to avoid estimating all entries
of the covariance matrix, but only directly those combinations
that are required -- as is, e.g., 
routinely done using the above test 
(\ref{SimpleDuan}). In the program, to be described later,
additional constraints to incorporate specifically accessible
measurement types (for example via interferometers, rather than
through homodyning measurements) can be taken into account 
via a finite number of linear constraints of the form
\begin{equation}
	\tr[R_i Z]=0,
\end{equation}
$i=1,\ldots,I$. If this set of $I$ constraints is too restrictive,
it may happen that no test is found that can detect the
entanglement.

\subsection{Remarks on quantitative statements}

One should be tempted to think that whenever a state
`violates' such a criterion by a large degree, it should be very much
entangled, in quantitative terms. For Gaussian states 
this is a simple issue, as the result
of the test is essentially just a lower bound to the logarithmic 
negativity, see above (compare also Ref.\ \cite{Anders}). 
More relevant, yet, are statements that do
not assume the Gaussian character of the state, as the very 
point of the test is that one not only does not need 
full tomographic knowledge 
of the quantum state, but not even knowledge of
all of its second moments. 

Yet, it is not true that the Gaussian state is the one with the 
smallest logarithmic negativity, given some value in Eq.\ 
(\ref{SimpleDuan}): 
there can be small violations. It has been shown that 
for a given value in Eq.\ (\ref{SimpleDuan}),
non-Gaussian
states may have a slightly smaller logarithmic 
negativity \cite{QuantRemark}. Another example has been
presented in Ref.\ \cite{New}, where it has been shown that 
for a fixed full covariance matrix (and not only a fixed 
value of Eq.\ (\ref{SimpleDuan})), there exists a non-Gaussian state
assuming a smaller negativity than the corresponding Gaussian state.
It is still true, however, that 
for a given left hand side of Eq.\ (\ref{SimpleDuan}), 
one can still find a lower bound of the 
logarithmic negativity, indicating
that `a state that very much violates this 
criterion is also very much entangled'.

For the {\it entanglement of
formation}, the smallest degree of entanglement 
is in turn assumed
for a Gaussian state for symmetric $1\times1$-mode states
\cite{GEOF}, 
as well as for the {\it squashed entanglement} \cite{New}. Also, 
the {\it conditional entropy}, 
\begin{equation}
	C(\rho)= S(\rho_A) - S(\rho)
\end{equation}	
for states $\rho$,
a lower bound to the distillable 
entanglement, takes in general 
its smallest value for Gaussian states \cite{Channel}.
Hence if the full covariance matrix of a 
state is known, then it is possible to 
evalute the entropies of the corresponding
Gaussian state with that covariance matrix,
yielding a lower bound to the distillable 
entanglement of the non-Gaussian state.
%
One hence does not have to assume the Gaussian character then:
One can then measures the moments, 
evaluates the entropies of the respective Gaussian
state, and take this value as lower bound to the
distillable entanglement of the true non-Gaussian state.


\section{Curved witnesses: all product criteria}
\label{sec:curved}

In this section, we now turn to curved witnesses, tests
that do not correspond to linear combinations of 
second moments, but to quadratic ones. Interestingly,
the measurements that have to be performed are
just the same ones as in linear tests, only the combination
of the respective outcomes is different. It turns out that
the use of quadratic tests is always advantageous to linear
tests. Frankly, it is always
 `better to multiply the numbers instead of adding them',
 see Figure 1.

Geometrically, such tests -- generalized product
criteria -- correspond to curved surfaces, which are
not hyperplanes. They are curved towards the
set of second moments of separable Gaussian states.
In the setting before, any witness $Z$ can be 
decomposed as
\begin{equation}
	Z= Z_x + Z_p + Z_{x,p},
\end{equation}
where 
\begin{eqnarray}
	P_x Z_x P_x &=& Z_x, \\
	P_p Z_p P_p &=& Z_p,\\
	P_x Z_p P_x &=& P_x Z_{x,p} P_x=0,\\ 
	 P_p Z_x P_p &=& P_p Z_{x,p} P_p = 0.
\end{eqnarray}	 
The projectors $P_x$ and $P_p$ are defined in Eq.\ (\ref{Px}) 
and (\ref{Pp}), respectively.
%
Let us first look at the class of witnesses $Z$ that are
of the form
\begin{equation}
	Z= Z_x + Z_p .
\end{equation}
From each witness of this form, a product criterion of the type of
Eq.\ (\ref{SimpleProduct}) can be derived,
which is stronger than the witness. Moreover, there is a one-to-one
correspondence between the witness and the product criterion.
Hence, all product tests in the strict sense, involving the
product of variances with respect to position and momentum
coordinates, are obtained. 

\begin{proposition}[All product criteria]
\label{prop:allprodcrit}
If $Z_x+Z_p$ is a witness, then 
\bea
	\label{eq:CWit1}
	\text{(i)} && \tr[Z_x\gamma_s]\,\tr[Z_p\gamma_s]\ge \frac{1}{4}\quad \text{for\ all\ separable\ $\gamma_s$,}\\
	\label{eq:CWit2}
	\text{(ii)} && \tr[Z_x\gamma]\,\tr[Z_p\gamma] < \frac{1}{4}\quad \text{for\ some\ entangled\ $\gamma$,}
\eea
and the set of entangled covariance matrices detected by this
quadratic test is strictly larger than that detected by the linear 
witness  $Z_x+Z_p$. In turn, if $Z_x$ and $Z_p$ are symmetric matrices
fulfilling $Z_x=P_x Z_x P_x$ and $Z_p=P_p Z_p P_p$ and the conditions
$(i)$ and $(ii)$, then $Z_x+Z_p$ is an entanglement
witness.
\end{proposition}

\newproof
This proof makes use of the material presented in the Appendix.
If the witness $Z_x+Z_p$ detects a covariance matrix $\gamma$,
	$\tr[(Z_x + Z_p  )\gamma]<1$,
then clearly also
	$\tr[Z_x \gamma]\, \tr[Z_p \gamma]<1/4$ holds.
However, if $\tr[(Z_x + Z_p)\gamma]\ge 1$, then
it does not follow directly that 
$\tr[Z_x \gamma]\, \tr[Z_p \gamma]\ge 1/4$.
Therefore, we continue by showing that if 
$\tr[Z_x \gamma]\, \tr[Z_p \gamma]<1/4$ then there 
exists a witness $Z'$ such that $\tr[Z'\gamma]<1$.
It follows that $\tr[Z_x \gamma]\, \tr[Z_p \gamma]<1/4$
for entangled covariance matrices $\gamma$ only,
concluding the proof of the statements (i) and (ii).

We define $Z'_x=a Z_x$ and $Z'_p=Z_x/a$, 
$a\in\rr\backslash\{0\}$, such that 
\begin{equation}	
	\tr[Z_x' \gamma] =  \tr[Z_p' \gamma].
\end{equation}
This, together with
$\tr[Z_x' \gamma]\, \tr[Z_p' \gamma]=\tr[Z_x \gamma]\, \tr[Z_p \gamma]<1/4$
implies that $\tr[(Z_x'+Z_p')\gamma]<1$.
Further, the matrix $Z'=Z_x'+Z_p'$ 
fulfils the witness condition
$\sum_k \text{str} Z'_k \ge 1/2$ from the Appendix,
where $Z'_k$ is the block on the diagonal of $Z'$ of party $k$:
since $Z'_k=S Z_k S^T$, where $S\in Sp(2n_k,\rr)$ is the 
symplectic transformation 
\begin{equation}
	S=(\sqrt{a}P_x+P_p/\sqrt{a}),
	\label{eq:Symtra}
\end{equation}	
the symplectic trace of $Z'_k$ is equal to that of $Z_k$.
Hence $Z'$ is a proper witness detecting $\gamma$.
Note that this argument is independent of whether $Z$
is a witness against full separability or for multi-party
entanglement.
	
Finally, the product criterion is detecting strictly 
more entangled covariance matrices 
than the witness $Z_x+Z_p$
since only the implication
$\tr[(Z_x+Z_p)\gamma]<1\Rightarrow \tr[Z_x\gamma]\,\tr[Z_p\gamma]<1/4$ holds,
while the converse 
direction does not necessarily hold 
for some entangled $\gamma$.
For instance, consider a witness $Z_x+Z_p$ 
and a covariance matrix $\gamma$ such that
\begin{equation}
\tr[(Z_x+Z_p)\gamma]<1.
\end{equation} 
The covariance matrices $S\gamma S^T$, where
$S=(\sqrt{a}P_x+P_p/\sqrt{a})$ as above, are proper covariance
matrices for all $a>0$. However, not all of them are detected by $Z_x+Z_p$,
since it is always possible to choose an $a>0$ such that
$\tr[(Z_x+Z_p)S\gamma S^T]=a\tr[Z_x\gamma]+\tr[Z_p\gamma]/a\ge 1$.
In contrast, the product witness detects the whole family since
\begin{equation}
	\tr[Z_x S\gamma S^T]\,\tr[Z_p S\gamma S^T]
	=\tr[Z_x\gamma]\,\tr[Z_p\gamma]<1/4.
\end{equation}	
Therefore, the product witnesses are stronger tests than the respective 
linear tests. 

Finally, assume that there exist symmetric matrices
$Z_x=P_x Z_x P_x$ and $Z_p=P_p Z_p P_p$, such that the conditions
$(i)$ and $(ii)$ are fulfilled. From condition $(i)$ it follows 
directly that $tr[(Z_x+Z_p)\gamma_s]\ge 1$ for separable $\gamma_s$. 
Then choose a $\gamma$ such that $(ii)$ holds. As before,
$\tr[Z_x\gamma(a)]\, \tr[Z_p\gamma(a)]=\tr[Z_x\gamma]\, \tr[Z_p\gamma]< 1/4$
for $\gamma(a)=S\gamma S^T$, where $S$ is the 
symplectic 
transformation~(\ref{eq:Symtra}). If we pick $a\in\rr\backslash\{0\}$
such that 
\begin{equation}
	\tr[Z_x\gamma(a)]=\tr[Z_p\gamma(a)], 
\end{equation}	
then it follows from
$\tr[Z_x\gamma(a)]\, \tr[Z_p\gamma(a)] < 1/4$ that
$\tr[(Z_x+Z_p)\gamma(a)] < 1$. Hence $Z_x+Z_p$ is an
entanglement witness.
\proofend

As a matter of fact, if we allow for $Z_{x,p}\neq 0$,
then we get a one-to-one correspondence of witnesses
and generalized product criteria, fully characterizing the convex set 
of separable second moments.

\begin{proposition}[All generalized product criteria]
If $Z=Z_x+Z_p+Z_{x,p}$ is a witness, then
\bea
	\label{eq:CGWit1}
	\text{(i)} && P_Z(\gamma_s)\ge \frac{1}{4}\quad \text{for\ all\ separable\ $\gamma_s$,}\\
	\label{eq:CGWit2}
	\text{(ii)} && P_Z(\gamma)< \frac{1}{4}\quad \text{for\ some\ entangled\ $\gamma$,}
\eea
where 
\begin{eqnarray}
	P_Z(\gamma) &=& \tr[Z_x\gamma]\,\tr[Z_p\gamma]\nonumber\\
	&+& \frac{1}{2}\tr[Z_{x,p}\gamma]-\frac{1}{4}(\tr[Z_{x,p}\gamma])^2,
\end{eqnarray}
and strictly more entangled covariances are detected than by
the witness $Z$. In turn, if $Z_x$, $Z_p$, and $Z_{x,p}$ are symmetric 
matrices fulfilling $Z_x=P_x Z_x P_x$, $Z_p=P_p Z_p P_p$, and $Z_{x,p}=P_x Z_{x,p} P_p+P_p Z_{x,p} P_x$
and the conditions $(i)$ and $(ii)$, then $Z_x+Z_p+Z_{x,p}$ is an entanglement
witness.
\end{proposition}

\newproof
If the witness $Z$ detects a covariance matrix $\gamma$,
$\tr[(Z_x + Z_p + Z_{x,p} )\gamma]<1$,
then also
\begin{equation}
	\tr[Z_x \gamma]\, \tr[Z_p \gamma]<A^2/4 
\end{equation}
holds,
where 
\begin{equation}
	A=1-\tr[Z_{x,p}\gamma], 
\end{equation}
	which is equivalent to
$P_Z(\gamma)<1/4$.
All the other steps of the proof of 
proposition \ref{prop:allprodcrit} can be performed in analogy, using
that $SZS^T=Z'_x+Z'_p+Z_{x,p}$, where 
$S=\sqrt{a}P_x+P_p/\sqrt{a}$ as above.
\proofend

These criteria hence form a complete set of criteria, and
all what has been said before is also applicable to these
tests. In a sense, these curved tests compensate for
local squeezings, operations under which the linear tests
are not invariant. Note also that the tests are quadratic, but
still in entries of the canonical coordinates. One can also
think of tests where the observables themselves include
higher polynomials. First interesting steps in this direction 
have been undertaken, for example, in Refs.\ \cite{C}.

\section{Numerical examples}
\label{sec:numerics}

We implemented the dual programs for witnesses detecting entanglement
and genuine multi-partite entanglement in Matlab (Version 7). The routines
have been made 
freely available \cite{ourprogs}. They make use of the solver
{\em SeDuMi} \cite{sedumi} and the interface 
{\em Yalmip} \cite{yalmip}, which
are also freely available.

\subsection{Testing full separability numerically}

The function {\tt FullyWit} implements the dual program 
of Eq.~(\ref{eq:dual1}). It is called by 
the line 
\jbox{\tt [c Z]=FullyWit(gamma,n,constraints).}
The inputs are the covariance matrix {\tt gamma}
and a vector {\tt n}, which holds the number of modes 
that each of the parties have. For instance, 
if {\tt gamma} is a 6-mode state held by three parties $A$, $B$, and $C$, 
where party $A$ holds 3 modes, party $B$ holds 1 mode, 
and party $C$ holds 2 modes, then {\tt n=[3 1 2]}.
The symmetric covariance matrix {\tt gamma} would 
have to have dimension $2n\times 2n$, where $n=6$.
Using the parameter {\tt constraints}, the witnesses
can be further restricted, as explained in Section \ref{sec:excon}.
Until then, we will set {\tt constraints=0}, thereby
not using this option.

The output {\tt Z} is a real symmetric matrix fulfilling
the first witness condition $\tr[\text{\tt Z}\gamma_s]\ge 1$ for all
separable covariances $\gamma_s$. 
The second output is 
\begin{equation}
	\text{{\tt c}}=\tr[\text{\tt Z\ gamma}]-1.
\end{equation}
Hence if {\tt c}$<0$, then {\tt gamma} is entangled,
and {\tt Z} is an optimal entanglement witnesses in
the sense of Proposition 1. Otherwise,
{\tt gamma} is separable.

All the following examples were solved in fractions
of a second on a Pentium 4 machine with $2.8$ GHz and 
$512$ Mb RAM by {\tt FullyWit} and {\tt MultiWit},
respectively.

The first example we would like to consider is the
PPT entangled state of $2\times 2$ modes given
in Ref.\ \cite{CVbe}
\begin{equation}
	\gamma_\textrm{WW}=
	\left[\begin{array}{ccccccccc}
		2 & 0 & 0 & 0 & 1 & 0 & 0 & 0\\
		0 & 1 & 0 & 0 & 0 & 0 & 0 & -1\\
		0 & 0 & 2 & 0 & 0 & 0 & -1&  0\\
		0 & 0 & 0 & 1 & 0 &-1 & 0 &  0\\
		1 & 0 & 0 & 0 & 2 & 0 & 0 &  0\\
		0 & 0 & 0 &-1 & 0 & 4 & 0 &  0\\
		0 & 0 &-1 & 0 & 0 & 0 & 2 &  0\\
		0 &-1 & 0 & 0 & 0 & 0 & 0 &  4
	\end{array}
	\right].
\end{equation}
The command 
\jbox{
{\tt [c Z]=FullyWit($\gamma_\textrm{WW}$,[2 2],0)}} 
yields
{\tt c=-0.1034}, exemplifying
the entanglement of the covariance matrix 
$\gamma_\textrm{WW}$, 
detected by the witness
\begin{equation}
	Z_\textrm{WW}=
	\left[\begin{array}{rrrrrrrr}
		x &  0 & 0 &  0 & -z &  0 & 0 &  0 \\
    0 &  2x & 0 &  0 & 0 &  0 & 0 &  z\\
   	0 &  0 &  x &  0 &  0 &  0 &  z & 0\\
    0 &  0 &  0 & 2x & 0 &  z &  0 & 0\\
   -z & 0 &  0 & 0 &   2y & 0 &  0 & 0\\
    0 & 	0 &  0 &  z & 0 &  y &  0 & 0\\
   	0 & 	0 &  z &  0 &  0 &  0 &  2y & 0\\
    0 &  	z & 0 & 0 & 0 & 0 & 0 &  y
	\end{array}
	\right],
\end{equation}
where $x=0.1394$, $y=0.0374$, and $z=0.1021$.

As a second example we consider the family
of GHZ like Gaussian states introduced in 
Ref.\ \cite{CVghz}. For $N$ modes,
the covariance matrix is given by
\begin{equation}
	\gamma_\textrm{GHZ}=
	\left[\begin{array}{ccccc}
		A&C&C &\ldots& C \\
		C&A& C & \ldots& C \\
		\vdots & & \ddots  & & \vdots\\ 
		&   &   & A & C\\
		C& C & \ldots & C & A\\
	\end{array}
	\right],
\end{equation}
where
\begin{equation}
	A= \left[
	\begin{array}{cc}
	a & 0 \\
	0 & b\\
	\end{array}
	\right], \,\,\,
	C= \left[
	\begin{array}{cc}
	c & 0 \\
	0 & d\\
	\end{array}
	\right],
\end{equation}
and
\begin{eqnarray}
	a&=&\frac{1}{N}e^{+2r_1}+\frac{N-1}{N}e^{-2r_2},\\
	b&=&\frac{1}{N}e^{-2r_1}+\frac{N-1}{N}e^{+2r_2},\\
	c&=&\frac{1}{N}(e^{+2r_1}-e^{-2r_2}),\\
	d&=&\frac{1}{N}(e^{-2r_1}-e^{+2r_2}).
\end{eqnarray}
Further, $r_1>0$ is the squeezing parameter of the first
initial mode at the beginning of the state construction
process, and $r_2>0$ is the squeezing parameter of the
other $N-1$ modes at that stage \cite{CVghz}.
These states have the following properties: 
they are pure, invariant under exchange of any two parties,
and are genuinely multi-partite entangled.

For $N=3$ and $r_1=r_2=(\ln 2)/2$, the covariance
matrix takes the simple form
\be
	\gamma_\textrm{3GHZ}=\frac{1}{2}
	\left[\begin{array}{ccccccc}
		  2 	&	0 	& 1 & 0 	& 1 & 0		\\
		  0 	& 3 &	0		& -1& 0 	&-1	\\
		  1 & 0 	&	2 	&	0 	&	1 &	0		\\
		  0 	& -1&	0 	&	3	& 0		& -1\\     
		  1 & 0 	&	1	& 0		& 2 	&	0		\\
		  0 	& -1& 0 	& -1& 0		&	3
	\end{array}\right].
	\label{eq:3GHZ}
\ee
The routine {\tt FullyWit} yields {\tt c=-0.500}, clearly 
demonstrating the entanglement of the state for these
parameters, and the witness is given by
\be
	Z_\textrm{fw}=
		\left[\begin{array}{ccccccc}
			2x 	& 0 	& -x 	& 0 	& -x 	& 0\\
			0		& 2x	& 0		&	2x 	& 0 	& 2x\\
			-x  & 0		& 2x	&	0 	&-x		& 0\\
			0		& 2x	& 0		&	2x 	& 0 	& 2x\\
			-x  & 0		& -x	&	0 	&2x		& 0\\
			0	&	2x	&	0 	&	2x 	&	0	& 2x
	\end{array}\right],
\ee
where $x=0.0833 \approx 1/12$. 

Note that the function {\tt FullyWit} can be used to 
perform all the tests necessary for the first 
classification of multiparty entanglement 
introduced in Section \ref{sec:classification}.
For instance, separability across the split $AB|C$ 
for three parties as in the last example can be tested by choosing 
\mbox{\tt n=[2 1]}. However, if the split $AC|B$ 
is chosen, then the parties $B$ and $C$ have to
be exchanged first by transforming $\gamma\mapsto S\gamma S^T$,
where
\be
	S=\left[\begin{array}{ccc}
			\Eins  	& 0 	& 0\\
			0 & 0	& \Eins \\
			0 &	\Eins & 0 
	\end{array}\right].
\ee
Then the test can be performed by the command 
\begin{equation}
\mbox{\tt [c Z]=FullyWit($S \gamma S^T$,[2 1],0)}.
\end{equation}
In a similar manner, all possible partitions for 
$N$ parties can be tested.

\subsection{Testing bi-separability numerically}

In analogy, the function {\tt MultiWit} 
implements the dual program of 
Eq.~(\ref{eq:Mdual1}). It is called by 
the line \jbox{\tt [c Z]=MultiWit(gamma,n,constraints).}
The inputs are again the covariance matrix {\tt gamma}
and a vector {\tt n}, which holds the number of modes 
that each of the parties have as in {\tt FullyWit}. 
Again, we put {\tt constraints=0}
and refer to Section \ref{sec:excon}.

The output {\tt Z} is a real symmetric matrix fulfilling
the first witness condition $\tr[\text{\tt Z}\gamma_s]\ge 1$ for all
bi-separable covariances $\gamma_s$. 
The second output is {\tt c}$=\tr[${\tt Z gamma}]-1.
Hence if {\tt c}$<0$, then {\tt gamma} is genuinely multi-partite
entangled, and {\tt Z} is an optimal entanglement 
witnesses in the sense of proposition 2. Otherwise,
{\tt gamma} is bi-separable.

Applying this routine to the 3 mode GHZ covariance of Eq.~(\ref{eq:3GHZ}),
we obtain {\tt c=-0.3056}, showing that the state is genuinely multi-partite
entangled. The witness detecting the state has the form
\be
	Z_\textrm{mw}=
		\left[\begin{array}{ccccccc}
			x 	& 0 	& -z 	& 0 	& -z 	& 0\\
			0		& y   & 0		&	w 	& 0 	& w\\
			-z  & 0		& x	&	0 	&-z		& 0\\
			0		& w	& 0		&	y 	& 0 	& w\\
			-z  & 0		& -z	&	0 	& x		& 0\\
			0	&	w	&	0 	&	w 	&	0	&  y
	\end{array}\right],
\ee
where $x=0.2315$, $y=0.2021$, $z=0.1157$, and $w=0.1875$.

Another example we consider is a four mode state occuring 
in an intermediate step of an continuous variable
entanglement swapping experiment as described in  
Refs.~\cite{Tan1999}. 
Key steps towards full implementations 
and first experimental implementations 
have been reported recently \cite{LeuchsNeu,Jia04}.

The idea can be briefly described as follows: 
two entangled states are produced between the modes 
labeled
$1$ and $2$ and between the modes $3$ and $4$, respectively.
Then a certain homodyne-type 
measurement is performed between
the modes $2$ and $3$. In a successful implementation, 
after this, entanglement can
be confirmed between the modes $1$ and $4$ which
have never interacted before. Here, we consider 
the state before
the actual measurement, which is the state 
where the modes $2$ and $3$ have been brought to 
overlap at a 50:50 beam splitter.

In order to see whether this can indeed be 
confirmed for realistic parameters, we construct
two mode entangled states with a fixed degree
of squeezing and a fixed degree of mixedness
by `backwards reasoning'. 
We start with the diagonal
matrix 
$\gamma'=\text{diag}(e^{-2r},e^{-2r},\alpha e^{2r},\alpha e^{2r})$,
where $r,\alpha > 0$.
This is the desired covariance matrix after
partial transposition and symplectic 
diagonalization. The entanglement is reflected
in the symplectic eigenvalue \cite{CV}
$e^{-2r}<1$ for $r>0$.
We arrive at the covariance matrix $\gamma$
by the reverse transformation, in this
case we first apply a 50:50
beam splitter and then partial transposition
with respect to the second system. The symplectic
eigenvalues of the resulting matrix $\gamma$ 
are both equal to $\sqrt{\alpha}$. The von-Neumann
entropy can for a Gaussian state 
be calculated as \cite{CV}
\be
	H(\rho)=\sum_{k=1}^n\left[ (N_k+1)\ln(N_k+1)-N_k\ln N_k\right],
\ee
where $N_k=(s_k-1)/2$ is obtained from
the symplectic eigenvalues $s_k$ of $\gamma$.

For both input entangled pairs of modes,
we choose $r=2\ln(2)/3$, corresponding to 4dB squeezing
of the initial state (the degree of squeezing can be read off
the smallest eigenvalue of the covariance matrix)
and $\alpha=5$, leading to
$H(\rho)=2.152$. The covariance matrix of the state 
after the beam splitter between the modes 2 and 3 
is then given by
\be
	\gamma_\text{SWAP}=\left[
	\begin{array}{ccccccccc}
			x 	& 0 	& y 	& 0 	& y & 0 & 0 & 0\\
			0 & x & 0 & -y & 0 & -y & 0 & 0\\
			y & 0 & x & 0 & 0 & 0 & y & 0\\
			0 & -y & 0 & x & 0 & 0 & 0 & -y\\
			y & 0 & 0 & 0 & x & 0 & -y & 0\\
			0 & -y & 0 & 0 & 0 & x & 0 & y\\
			0 & 0 & y & 0 & -y & 0 & x & 0 \\
			0 & 0 & 0 & -y & 0 & y & 0 & x
	\end{array}
	\right],
	\label{eq:SWAP}
\ee
where $x=6.4980$ and $y=-4.3142$.
The routine {\tt MultiWit} returned {\tt c=-0.2305}, demonstrating 
the genuine four-partite 
entanglement of the state. The corresponding
witness is of the form of Eq.~(\ref{eq:SWAP}), where now
$x=0.2352$ and $y=0.1660$. 

The function {\tt FullyWit} returns
{\tt c=-0.6031} for {\tt n=[1 1 1 1]}. 
The corresponding witness is 
of the same form as Eq.~(\ref{eq:SWAP}),
where now $x=0.125$ and $y=0.0884$.
However, this specific instance of a test 
can only clarify that the 
state is not fully separable.

\subsection{Further experimental constraints}
\label{sec:excon}

If the experimental set up limits the 
possible set of tomographic measurements, then
these constraints can be taken into account
by requiring that the witness $Z$ fulfils
\be 
 	\tr[Z A]=0,
 	\label{eq:addcon}
\ee 
where $A$ is an operator describing the measurement.
We can, for instance, set the element $Z_{j,k}$ to zero
with the matrix $A=F_{j,k}$ defined in 
Eq.~(\ref{eq:sbasis}). In this way it is possible 
to restrict $Z$ such that no position-momentum
correlation measurements are required, as discussed
in Sec. \ref{sec:bd_xp}.

If we require that Eq.~(\ref{eq:addcon}) holds, then
the only constraint that is affected in the
primal program is 
\be
	\gamma\ge\gamma_A\oplus\gamma_B\ 
	\mapsto\ \gamma+x_A A\ge\gamma_A\oplus\gamma_B,
\ee
and in analogy for the program detecting only genuine 
multi-partite entanglement. Here, $x_A$ is a new
real variable in addition to the ones collected in
the real vector $x$ of the primal program.
Hence, the effect of the
additional constraint~(\ref{eq:addcon}) in the dual
program on the primal program is, that the set of
separable states is effectively enlarged. This was
to be expected, since the set of witnesses
is further restricted, and hence less entangled
states can be detected with further constraints.

If the dual program finds a restricted 
witness $Z_r$ such that $\tr[Z_r\gamma]<1$ for
the covariance $\gamma$ in question, then  
the experimental proof of entanglement can be
simplified, otherwise some or all of 
the additional constraints have to be dropped.
In the program, for each constraint a matrix {\tt A}
has to be defined of the dimensions of the covariance
matrix. As already mentioned, if no constraint is desired,
then {\tt constraints=0}. Otherwise, 
\jbox{{\tt constraints=[A1 A2 ... Ak]},}
when {\tt k} extra constraints of the form of 
Eq.~(\ref{eq:addcon}) are included. 

\section{Summary}

In this paper, we have provided a complete
picture of all 
tests for continuous-variable entanglement which are
linear in variances of canonical coordinates as solutions of 
certain semi-definite problems. This framework has turned
out to be applicable both in the bi-partite case, as well as for 
the various separability classes in the multi-partite setting.
We moreover classified all product criteria, leading to 
curved witnesses, so non-linear witnesses,
curved towards the set of separable covariance 
matrices. 

Finally, we presented the functioning of the two
routines  {\tt FullyWit} and {\tt MultiWit}, which deliver just
such optimal tests, given an assumption on how the state
should roughly be like. We discussed several examples in
detail. It is the hope that this picture, and also the freely 
avaliable routines, provide useful practical tools in assessing
entanglement in continuous-variable systems, 
also in the experimental
context.

\section{Acknowledgements}

We would like to thank 
J.\ Anders, 
M.\ Aspelmeyer,
U.L.\ Andersen,
F.\ Brandao,
C.\ Brukner,
J.I.\ Cirac,
J.\ Fiurasek,
G.\ Giedke,
O.\ Gl\"ockl,
O.\ G{\"u}hne,
G.\ Leuchs,
M.\ Lewenstein,
P.\ van Loock,
M.B.\ Plenio,
O.\ Pfister,
R.\ Schnabel,
C.\ Silberhorn,
R.F.\ Werner, and
M.M.\ Wolf	
for discussions on the subject of this paper. We would especially
like to thank J.\ L{\"o}fberg for expansive communication
on structure and functioning 
of primal-dual solvers, and very
significant help in optimizing the software.
This work has been supported 
by the DFG (SPP 1116, SPP 1078),
the EU (QUPRODIS, QAP), the EPSRC, and the
European Research Councils (EURYI).

\section*{Appendix: Classification of all entanglement witnesses}

Unlike the case of entanglement witnesses based on states,
all separating hyperplanes of the set of fully separable covariance
matrices can be very clearly characterized for a general $n$-mode 
system. This is presented in Ref.\ \cite{GBook}, and we briefly
state this result here without proof for completeness only:

\begin{theorem}
[{\rm \cite{GBook}}]
$Z$ is an entanglement witness based on second moments
in the sense of (\ref{eq:FWit1}) and (\ref{eq:FWit2}) 
if and only if
\begin{eqnarray}
	\label{eq:witness1}
	(i) && Z \ge 0, \\
	\label{eq:witness2}
	(ii) && \sum_{k=1}^{n}\str[Z_k] \ge \frac{1}{2},\\
	\label{eq:witness3}
	(iii) && \str[Z] < \frac{1}{2}
\end{eqnarray}
holds, 
where $Z_k$ is the block on the diagonal of $Z$ 
acting on system labeled $k$.
\end{theorem}
This characterizes simply all linear 
entanglement witnesses based on second moments. 
Further, the set of bi-separable covariance matrices
is also convex and closed, since if a state contains
an (arbitrary small) multi-partite entangled part
in every decomposition, then all the states in
its neighborhood will also contain such a part.
It follows that entanglement witnesses
can be constructed that detect only genuine multi-partite
entangled states. The conditions for such witnesses
are \cite{GBook}
\bea
	\label{eq:MWit1}
	(i) && Z \ge 0, \\
	\label{eq:MWit2}
	(ii) && \sum_{j=1}^{M}\str[Z_{\pi(k)}^{(j)}] \ge \frac{1}{2}\quad\textrm{for\ all\ }k,\\
	\label{eq:MWit3}
	(iii) && \str[Z] < \frac{1}{2},
\eea
where $\pi(k)$ is a partition of the $n$ modes into $M<n$ 
parties as above, and $Z_{\pi(k)}^{(j)}$ is the block on the diagonal of $Z$ 
of the $j$-th party of the partition $\pi(k)$.
For excluding bi-separable states it is sufficient to 
consider partitions into just $M=2$ parties.


\begin{thebibliography}{99}
\bibitem{Remark}
	This has been folk knowledge in the field since quite a 
	while. To one of the authors it has first pointed out to 
	by F.\ Verstraete.
\end{thebibliography}
\end{document}